\documentclass[iop]{emulateapj}
\usepackage{color}
\usepackage{graphicx}
\usepackage{amsmath}

\shorttitle{Testing the cosmic opacity with strongly lensed
supernova Ia} \shortauthors{Ma et al.}

\begin{document}
\title{Testing cosmic opacity with the combination of strongly lensed and unlensed supernova Ia }

\author{Yu-Bo Ma\altaffilmark{1}, Shuo Cao\altaffilmark{2$\ast$}, Jia Zhang\altaffilmark{3}, Jingzhao Qi\altaffilmark{4$\dag$}, Tonghua Liu\altaffilmark{2}, Yuting Liu\altaffilmark{2}, and Shuaibo Geng\altaffilmark{2}}

\altaffiltext{1}{Department of Physics, Shanxi Datong University,
Datong, 037009, China;} \altaffiltext{2}{Department of Astronomy,
Beijing Normal University, Beijing 100875, China;
\emph{caoshuo@bnu.edu.cn}}\altaffiltext{3}{School of physics and
Electrical Engineering, Weinan Normal University, Shanxi 714099,
China;} \altaffiltext{4}{Department of Physics, College of Sciences,
Northeastern University, Shenyang 110819, China;
\emph{qijingzhao@mail.neu.edu.cn}}

\begin{abstract}

In this paper, we present a scheme to investigate the opacity of the
Universe in a cosmological-model-independent way, with the
combination of current and future measurements of type Ia supernova
sample and galactic-scale strong gravitational lensing systems with
SNe Ia acting as background sources. The observational data include
the current newly-compiled SNe Ia data (Pantheon sample) and
simulated sample of SNe Ia observed by the forthcoming Large
Synoptic Survey Telescope (LSST) survey, which are taken for
luminosity distances ($D_\mathrm{L}$) possibly affected by the
cosmic opacity, as well as strongly lensed SNe Ia observed by the
LSST, which are responsible for providing the observed time-delay
distance ($D_\mathrm{\Delta t}$) unaffected by the cosmic opacity.
Two parameterizations, $\tau(z)=2\beta z$ and
$\tau(z)=(1+z)^{2\beta}-1$ are adopted for the optical depth
associated to the cosmic absorption. Focusing on only one specific
type of standard cosmological probe, this provides an original
method to measure cosmic opacity at high precision. Working on the
simulated sample of strongly lensed SNe Ia observed by the LSST in
10 year $z$-band search, our results show that, with the combination
of the current newly-compiled SNe Ia data (Pantheon sample), there
is no significant deviation from the transparency of the Universe at
the current observational data level. Moreover, strongly lensed SNe
Ia in a 10 year LSST $z$-band search would produce more robust
constraints on the validity of cosmic transparency (at the precision
of $\Delta\beta=10^{-2}$), with a larger sample of unlensed SNe Ia
detected in future LSST survey. We have also discussed the ways in
which our methodology could be improved, with the combination of
current and future available data in gravitational wave (GW) and
electromagnetic (EM) domain. Therefore, the proposed method will
allow not only to check the foundations of observational cosmology
(a transparent universe), but also open the way to identify
completely new physics (non-standard physics).

\end{abstract}


\maketitle


\section{Introduction}

The discovery of cosmic acceleration is arguably one of the most
important developments in modern cosmology, which is supported by
the fact that type Ia supernovae (SNe Ia) are observed to be fainter
than expected in a decelerating universe
\citep{Riess98,Perlmutter99}. With the inclusion of a mysterious
component with negative pressure as a new cosmological component, a
large number of dark energy models have been proposed to explain the
cosmic acceleration
\citep{Ratra88,Caldwell98,Cao11a,Cao13a,Cao14,Ma17,Qi18}. There are,
however, another theoretical approaches trying to explain cosmic
acceleration by modification of gravity at cosmological scales
\citep{Qi17,Xu18}. On the other hand, without introducing the new
component in the Universe, some popular theories of there had been
debates on the interpretation of underlying physical mechanism for
the observed SNe Ia dimming. Some popular theories include the
absorption, scattering or axion-photon mixing due to the dust in our
galaxy \citep{Tolman30}, and possible oscillation of photons
propagating in extragalactic magnetic fields
\citep{Aguirre99,Csaki02}. In this paper, we focus on the former
case, in which the deviation of photon number conservation is
related to the correction of Tolman test, equivalent to measurements
of the well-known distance-duality relation (DDR)
\citep{Etherington1,Etherington2,Cao11}
\begin{equation}
\frac{D_\mathrm{L}}{D_\mathrm{A}}(1+z)^{-2}=1,
\end{equation}
where $D_\mathrm{L}$ and $D_\mathrm{A}$ are respectively the
luminosity distance (LD) and the angular diameter distance (ADD) at
the same redshift $z$. The DDR holds when the geodesic deviation
equation is valid, photons follow null geodesic, and the number of
photon is conserved \citep{Ellis}. Thus, the possibilities of the
DDR violation are: evidence for a non-metric theory of gravity in
which photons do not follow null geodesic, and non-conservation of
the number of photons. If one considers that the photon traveling
along null geodesic is more fundamental and unassailable, the
violation of DDR most likely implies non-conservation of the photon
number, which can be related to presence of some opacity source
(gravitational lensing and dust extinction) and non-standard exotic
physics \citep{Bassett04,Corasaniti06}. Therefore, it is rewarding
to explore the DDR to test the validity of photon conservation and
related phenomena. In this case, the flux received by the observer
will be reduced by a factor $e^{\tau(z)/2}$, and observed luminosity
distance can be obtained by
\begin{equation}{\label{eq3}}
D_{\mathrm{L,obs}}=D_{\mathrm{L,true}}\cdot e^{\tau/2}\;,
\end{equation}
where $\tau$ is the opacity parameter which denotes the optical
depth associated to the cosmic absorption. Testing this quality with
high accuracy can also provide a powerful probe of the transparency
of the Universe.

Several tests have been proposed in the past years assuming a opaque
universe. The original idea of studying the cosmic opacity in the
framework of a flat $\Lambda$CDM model can be traced back to
\citet{More09}, which examined the difference of the opacity
parameter at redshifts $z=0.20$ and $z=0.35$, from two sub-samples
of ESSENCE SNe Ia \citep{Davis07} and the corresponding distance
measurements of BAO as a standard ruler in the radial direction
\citep{Percival07}. Further papers have also used the measurements
of the cosmic expansion $H(z)$ from cosmic chronometers to place
constraints on the matter density parameter $\Omega_m$, and
investigated the cosmic opacity in flat $\Lambda$CDM model
\citep{Avgoustidis10}. While comparing the results from the
luminosity distance with those obtained from the Union SNe
compilation data \citep{Kowalski08}, differences in central values
of the best-fit cosmic opacity parameters were also reported:
$\Delta \tau <0.012$ (95\% C. L.) for the redshift range between 0.2
and 0.35. However, it should be noted that all these studies
concerning the cosmic opacity are still model-dependent. By means of
astronomical observations, \citet{Holanda12} proposed a
model-independent estimate of $D_L$ which are obtained from a
numerical integration of $H(z)$ data and then confronted with the
observed one from SNe Ia observations. Such methodology was then
extended with three model-independent methods, which used SNe Ia
data to get the luminosity distances at the redshifts corresponding
to $H(z)$ data through interpolation method, smoothing method and
nearby SNe Ia method \citep{Liao13}. A better way to observationally
test the cosmic opacity is via independent measurements of intrinsic
luminosities and sizes of the same object, without using a specific
cosmological model. It is well known that type Ia supernovae are the
ideal tool to estimate the luminosity distances, while the angular
diameter distances are derived from various astrophysical probes
including the Sunyaev-Zeldovich effect together with x-ray emission
of galaxy clusters \citep{Filippis05,Bonamente06,Cao16b}, as well as
the gas mass fraction in galaxy clusters \citep{Allen08}. For
instance, the analysis performed by \citet{Li13} has revealed that a
transparent universe is ruled out by the \citet{Bonamente06} sample
at 68.3\% confidence level (C.L.), which demonstrated the importance
of considering the dimming effect of the largest Union2.1 SNe Ia
sample \citep{Suzuki12}. However, given the limited sample size of
current ADD measurements, one has to take care of the errors due to
the mismatch between the ADD redshift and the closest SNe Ia in the
companion SNe Ia sample adopted. In addition, other attempts focused
on the gravitational wave signal from inspiraling binary system to
determine the absolute value of their luminosity distances in the
redshift range of $0<z<5.0$. More importantly, in the framework of
FLRW metric, GWs propagate freely through a perfect fluid without
any absorption and dissipation. Therefore, when confronting the
luminosity distance derived from SNe Ia with that directly measured
from GW sources, one may naturally propose a scheme to investigate
the opacity of the Universe. Such original proposal, based on the
simulated data of gravitational waves from the third-generation
gravitational wave detector (the Einstein Telescope, ET), has been
extensively discussed in the recent works of \citet{Qi19b,Wei19}.
The result -- first prediction of the comic-opacity measurement
using GWs -- confirmed the accurate constraints on the cosmic
opacity. However, in order to place stringent constraints on the
cosmic opacity, we need the $D_L(z)$ measurements both from SNe Ia
and GW at the same redshift. Therefore, the redshift mismatch
between the GW events and the SNe Ia sample, due to the so-called
``redshift desert" problem still remains challenging with respect to
the exploration of the cosmic opacity.

Due to astrophysical complications and instrumental limitations, it
is difficult to observe both the luminosity distance and the angular
diameter distance of the same source simultaneously. In this
context, it is clear that collection of more complete observational
data concerning the multiple measurements of the same type of
astrophysical probe does play a crucial role. In this paper, we
focus on the combination of current and future measurements of type
Ia supernova sample and galactic-scale strong gravitational lensing
systems with SNe Ia acting as background sources. More specifically,
a completely model independent approach will be used to constrain
cosmic opacity using the time-delay observations of strong
gravitational lensing systems as standard rulers. Based on a
reliable knowledge about the lensing system, i.e., the Einstein
radius (from image astrometry) and stellar velocity dispersion (form
central velocity dispersion obtained from spectroscopy), one can use
it to derive the information of ADDs
\citep{Grillo08,Biesiada10,Cao12a,Cao12b,Cao12c,Cao13,Li16,Cao17c,Ma19},
test the weak-field metric on kiloparsec scales
\citep{Cao15,Collett18}, and probe the distance duality relation in
a cosmological model independent approach \citep{Liao16,Yang19}. In
addition, multiple images of the lensed variable sources take
different time to complete their travel and the time delay is a
function of the Fermat potential difference, and three angular
diameter distances between the observer, lens, and source
\citep{Treu10}. Therefore, strong-lensing time delays between
multiple images ($\Delta t$) can be used to derive the the so-called
time-delay distance ($D_\mathrm{\Delta t}$) unaffected by the cosmic
opacity. It is of interest to note that time delay between the
images of strongly lensed SNe Ia provide a laboratory to probe such
possibility \citep{Refsdal64}. More specifically, due to
exceptionally well-characterized spectral sequences and relatively
small variation in quickly evolving light curve shapes and color
\citep{Nugent02,Pereira13}, strongly lensed SNe Ia (SLSNe Ia) have
notable advantages over traditional strong lenses as time-delay
indicators (AGNs and quasars). Recently, the measurements of time
delay distance between multiple images of lensed SNe Ia have become
an effective probe in cosmology, which opens a possibility to test
the speed of light on the baseline up to the redshift of the source
\citep{Cao18}, as well the validity of the FLRW metric
\citep{Qi19a}. It is well known that the discovery of a new
gravitationally lensed type Ia supernova (SNe Ia) iPTF16geu (SN
2016geu) from the intermediate Palomar Transient Factory (iPTF) has
opened up a wide range of possibilities of using strong lensing
systems in cosmology and astrophysics \citep{Goobar17}. Strong
lensing time-delay predictions for this system were discussed in
detail in \citet{More17}. Focusing on the forthcoming Large Synoptic
Survey Telescope (LSST) survey, \citet{Goldstein17} have made a
detailed calculation of the lensing rate caused by lensing galaxies,
with the final results showing that at its design sensitivity LSST
would register about 650 multiply imaged SNe Ia in a 10 year
$z$-band search. The purpose of our analysis is to show how the
significantly improved measurements of galactic-scale strong
gravitational lensing systems with SNe Ia acting as background
sources can be used to probe the opacity of the Universe.

In order to compare our results with the previous one of
\citet{Li13,Liao13}, in our analysis we consider two particular
parameterizations of phenomenological $\tau(z)$ dependence: (i)
$\tau(z)=2\beta z$; (ii) $\tau(z)=(1+z)^{2\beta}-1$ to describe the
optical depth associated to the cosmic absorption. This paper is
organized as follows. We introduce the methodology in Section 2,
while the current SNe Ia data and simulated SLSNe Ia data used in
our work are presented in Section 3. The statistical method and
constraint results on cosmic opacity parameters are illustrated in
Section 4. Finally, we summarize our main conclusions and make a
discussion in Section 5.

\section{Methodology}

In order to measure the luminosity distance, we always turn to
luminous sources of known (or standardizable) intrinsic luminosity
in the Universe, such as SNe Ia in the role of standard candles.
However, it should be emphasized that the cosmic absorption could
affect the luminosity distance measurements of SNe Ia observations
as shown in Eq.~(2). More specifically, if the universe is opaque,
the flux from SNe Ia received by the observer will be reduced, and a
straightforward solution is to characterize this effect with a
factor $e^{-\tau(z)}$, where $\tau(z)$ is the optical depth related
to the cosmic absorption. From this point of view, from the
information of the luminosity distance for each SNe Ia provided by
the current and future SNe Ia surveys, one can be directly derive
the corresponding luminosity distance in an opaque Universe, which
can be finally used to test the transparency of the Universe.

On the other hand, in this paper, the angular diameter distances are
obtained from SLSNe Ia observations in a cosmological-model
independent way. As one of the successful predictions of general
relativity in the past decades, strong gravitational lensing has
become a very important astrophysical tool \citep{Walsh79,Young81},
allowing us to use lensing galaxies with distorted multiple images
of the background sources to act as time-delay indicators
\citep{Suyu13,Suyu14}. For a specific strong-lensing system with the
lensing galaxy at redshift $z_l$ and lensed SNe Ia at redshift
$z_s$, the angular diameter distance in a spatially at FLRW universe
\begin{eqnarray}
\label{inted} D_A(z_1, z_2)=\frac{c_{z_s}}{H_{0}
(1+z_2)}\int_{z_1}^{z_2} \frac{dz'}{E(z')} \, ,
\end{eqnarray}
can be directly derived from the so-called time delay distance
\begin{equation}
D_{\mathrm{\Delta t}}=\frac{D_{\mathrm{l}}
D_{\mathrm{s}}}{D_{\mathrm{ls}}}.
\end{equation}
where $D_{ls}$ and $D_s$ are angular diameter distances between the
lens and the source and between the observer and the source. Here
$E(z)$ is the dimensionless Hubble parameter and $H_0$ is the Hubble
constant. The light from each image in a lensing system takes a
different path through the lens before reaching the observer. If the
lensed object is a variable source, the images vary asynchronously
with a geometrical time delay based on these path differences.
Considering the fact that the background source SNe Ia is a
transient event with well defined light curve after peak, strong
gravitational time delays between the multiple images will be
revealed in the photometry, due to different light paths combined
with the well-known Shapiro effect \citep{Schneider92}. More
specifically, the time delay distance $D_{\mathrm{\Delta t}}$ and
the lensing system observations can be linked by the following
formula
\begin{equation}
\Delta t_{i,j} = \frac{D_{\mathrm{\Delta
t}}(1+z_{\mathrm{l}})}{c_{z_s}}\Delta \phi_{i,j}, \label{relation}
\end{equation}
where $\Delta t_{i,j}$ is the time delay between images of the
lensed source obtained from lensed SNe Ia, and $\Delta\phi_{i,j}$ is
the Fermat potential difference between image positions
\begin{equation}
\Delta\phi_{i,j}=[(\boldsymbol{\theta}_i-\boldsymbol{\beta})^2/2-\psi(\boldsymbol{\theta}_i)-(\boldsymbol{\theta}_j-\boldsymbol{\beta})^2/2+\psi(\boldsymbol{\theta}_j)]
\end{equation}
where $\boldsymbol{\theta}_i$ and $\boldsymbol{\theta}_j$ represent
the position of images of the lensed source, $\beta$ is the source
position and $\psi$ denotes two-dimensional lensing potential
related to the mass distribution of the lens. Therefore, the
unaffected time-delay distance can be precisely derived from the
observations of strongly lensed SNe Ia. In actual calculations,
given the relation between the angular diameter distance $D_A$ to
the proper distance $D_P$ in the flat FLRW metric, the distance
ratio of $D_{ls}/D_s$ can be expressed in terms of angular diameter
distances $D_{l}$ and $D_s$
\begin{equation}\label{ratio}
 \frac{D_{ls}}{D_s}=1-\frac{1+z_l}{1+z_s}\frac{D_l}{D_s}
\end{equation}
Now, considering the DDR with the following form used in our
analysis and in the literature \citep{Li13,Liao13}
\begin{equation}
D_\mathrm{A}=\frac{D_\mathrm{L,obs}}{(1+z)^{2}}e^{-\tau/2},
\end{equation}
the observational counterpart of the time-delay distance with the
cosmic opacity $\tau$ can be determined from the angular diameter
distances $D_l$ and $D_s$ by fitting to unlensed SNe Ia. We will use
the current observations of the Pantheon sample consisting of 1048
unlensed SN Ia \citep{Scolnic18}, as well as the simulated unlensed
and lensed SNe Ia data from LSST to test, model-independently, the
possible violation of the DDR, which can be translated to possible
existence of cosmic opacity.

\begin{table*}
\begin{center}
\begin{tabular}{lcccc}
\hline\hline

& $\delta\Delta t$ & $\delta\Delta t (ML)$ & $\delta\Delta\psi$ &  $\delta\Delta\psi (LOS)$  \\
\hline
SLSNe Ia  &1\% & \citet{Pierel19} & 3\% &1\% \\
\hline & $\sigma_{\rm mean}$  & $\sigma_{\rm int}$ & $\sigma_{\rm lens}$ & $\sigma_{sys}$ \\
\hline
SNe Ia & 0.08 mag  & 0.09 mag & 0.07$z$ mag & $0.01(1+z)/1.8$ mag\\
\hline\hline
\end{tabular}
\end{center}
\caption{The relative/ absolute uncertainties of factors
contributing to the distance measurements for the lensed and
unlensed SNe Ia sample. $\delta\Delta t (ML)$ and $\delta LOS$
correspond to macrolensing effect and and light-of-sight
contamination, respectively. }\label{error}
\end{table*}

\section{Observations and simulations}

In the following, we describe the data sets that we will use in the
present analysis.

\subsection{Current observations of unlensed SNe Ia}

First of all, for the current observations of SNe Ia, we use the
recent Pantheon compilation of 1048 SNe Ia released by the
Pan-STARRS1 (PS1) Medium Deep Survey \citep{Scolnic18}. Covering the
redshift range $0.01<z<2.3$, the observed distance modulus of each
SNe is given by
\begin{eqnarray}
\mu _ { \mathrm { SN } }&=& m_{B}+\alpha \cdot X _ { 1 } - \beta
\cdot \mathcal{C} - M _ { B },
\end{eqnarray}
where $m_{B}$ is the rest frame \textit{B}-band peak magnitude,
$M_{B}$ is the absolute \textit{B}-band magnitude, $X_1$ and
$\mathcal{C}$ describe the time stretch of light curve and the
supernova color at maximum brightness, respectively. Note that
$m_B$, $X_{1}$, and $\mathcal{C}$ can be obtained from the observed
SNe light-curves, while there are always three nuisance parameters
($\alpha$, $\beta$, and $M_{B}$) to be fitted in the distance
estimate. To dodge this problem, based on the approach proposed by
\citet{Marriner11} and including extensive simulations for
correcting the SALT2 light curve fitter, \citet{Kessler17} proposed
a new method called BEAMS with Bias Corrections (BBC) to calibrated
each SNe Ia. Therefore, for the Pantheon sample, the
stretch-luminosity parameter $\alpha$ and the color-luminosity
parameter $\beta$ is calibrated to zero, and the observed distance
modulus is simply reduced to $\mu_{\rm{SN}}= m_B-M_B$
\citep{Scolnic18}. Therefore, for each SNe Ia, the luminosity
distances $D_L(z)$ considering the possible effect of cosmic opacity
can be calculated from the definition of
\begin{equation}
D_L(z)=10^{(m_{B}-M_B)/5-5} (Mpc).
\end{equation}
which will be used to provide the measurement of the
opacity-dependent luminosity distance. Please refer to
\citet{Scolnic18} for detailed information of the Pantheon SNe Ia
sample, which has been widely applied to place stringent constraints
on the cosmological parameters \citep{Qi18,Qi19b}.

\subsection{Simulated data of lensed and unlensed SNe Ia}

Following the recent detailed calculation of the likely yields of
several planned strong lensing surveys based on realistic simulation
of lenses and sources \citep{Goldstein17}, it was predicted that
$\sim$ 930 SLSNe Ia will be discovered by the Large Synoptic Survey
Telescope (LSST) over its 10 year survey, with 70\% of the SLSNe Ia
having time delays that can be measured precisely. Therefore, LSST
can increase the detection of lensed SNe Ia by an order of magnitude
and yield 650 multiply imaged SNe Ia in a 10 year $z$-band search.
Moreover, it was reveled that $10^6$ unlensed type-Ia supernovae
candidates are expected to be identified in cadenced, wide-field
optical imaging surveys including LSST \citep{Cullan17}. Next we
simulate a set of SLSNe Ia/SNe Ia events. The standard $\Lambda$CDM
model is taken as our transparent cosmological model with fiducial
values $\Omega_m=0.308$, $H_0=67.8$ km/s/Mpc from the current Planck
2015 data \citep{Ade15}.

For the strongly lensed SNe Ia, we carry out a Monte Carlo
simulation of the lens and source populations to forecast the yields
of multiply imaged SNe Ia for LSST. The specific steps to simulate
the mock data are listed as follows, which is similar with that used
in \citet{Cao18}:

I. In this analysis, we consider only the strong gravitational
lensing of SNe Ia by early-type galaxies, the velocity distribution
of which is modeled as a modified Schechter function with parameters
from the SDSS DR3 data \citep{Choi07}£º
\begin{equation}
\frac{d n}{d \sigma}= n_*\left(
\frac{\sigma}{\sigma_*}\right)^\alpha \exp \left[ -\left(
\frac{\sigma}{\sigma_*}\right)^\beta\right] \frac{\beta}{\Gamma
(\alpha/\beta)} \frac{1}{\sigma} \, , \end{equation} where $\alpha$
is the low-velocity power-law index, $\beta$ is the high-velocity
exponential cut-off index, $n_*$ is the integrated number density of
galaxies, and $\sigma_*$ is the characteristic velocity dispersion.
Such function also quantifies the sampling distribution (redshift
distribution) of the galactic-scale lenses. We model the mass
distribution of the lens galaxies as a singular isothermal ellipsoid
(SIE), which is accurate enough as first-order approximations to the
mean properties of galaxies relevant to statistical lensing
\citep{Koopmans09,More16,Cao16a}. In this model, the Einstein radius
is given by
\begin{equation}
\theta_{E} = \lambda(e) 4\pi
\left(\frac{\sigma}{c}\right)^2\frac{D_{\rm ls}}{D_{\rm s}}
\end{equation}
where $\sigma$ is the velocity dispersion of the lens galaxy, and
$e$ is its ellipticity. In our fiducial model, the three dimensional
shapes of lens galaxies are characterized in the combination of two
equal number of extreme case \citep{Chae03}, while the so-called
``dynamical normalization" $\Lambda(e)$ is related to the lens
ellipticity, the distribution of which is modeled as a Gaussian
distribution with $e=0.3\pm0.16$ \citep{Oguri08}. Using the
simulation programs publicly available \citep{Collett15}, we obtain
a population of strong lensing systems on the base of realistic
population models. The population of strong lenses is dominated by
galaxies with velocity dispersion of $\sigma_0=210\pm50$ km/s, while
the lens redshift distribution is well approximated by a Gaussian
distribution with mean $z_l=0.80$. These results are well consistent
with what the LSST survey might yield in the future
\citep{Collett15}. The redshift distribution of the multiply imaged
SNe Ia takes the form of \citet{Goldstein17}, which furthermore
constituted the differential rates of lensed SNe Ia events as a
function of source redshift $z_s$ in a 10 year LSST z-band search.
In each simulation, there are 650 type Ia supernova covering the
redshift range of $0<z_s<1.70$. The source position is randomly
sampled within the Einstein radius ($\theta_E$) at the source plane.

II. We derive the time delays for each system from Eqs.~(5)-(6),
which depend on the measurements of redshifts, lens velocity
dispersion, Fermat potential difference between two image positions
and the relative source position on the source plane. For each
SNeIa-galaxy lensing system, the redshifts of the lens
$z_{\mathrm{l}}$ and the source $z_{\mathrm{s}}$ can be precisely
measured at the current observational level, while three key
ingredients (stellar velocity dispersion, high-resolution images of
the lensing systems, and time delays) can be derived concerning the
imaging and spectroscopy from the Hubble Space Telescope (HST) and
ground-based observatories. On the one hand, benefit from the
state-of-the-art lens modelling techniques \citep{Suyu10,Suyu12b}
and kinematic modelling methods \citep{Auger10,Sonnenfeld12}, one
can place stringent limits on the image positions, the source
position, and the Einstein radius from current high-resolution image
astrometry.

III. On the other hand, concerning the strategy of error estimation,
three sources of uncertainties are included in our simulation.
Firstly, due to exceptionally well-characterized spectral sequences
and considerable variation in light curve morphology
\citep{Nugent02,Pereira13}, the time delay between each image can be
precisely measured from the time-domain information observed by
dedicated monitoring campaigns. More interestingly, SLSNe Ia time
delays can be obtained in a single observing season, since the light
curves have a strong peak before they decay, occurring over a
time-scale of several weeks \citep{Foxley-Marrable17}. In the
framework of a typical SNe Ia-elliptical galaxy lensing systems, the
fractional uncertainty of $\Delta t$ is expected to be determined at
the level of 1\%, which is supported by the recent analysis by the
strong lens time delay challenge (TDC) \citep{Liao2015,Greg2015}.
Secondly, despite of these advantages, lensed SNe Ia still face the
problem due to the effect of microlensing by stars in the lens
galaxy. More progress has been made to discuss this important issue
\citep{Goldstein17}, which indicated that the absolute time delay
error due to microlensing is unbiased at the sub-percent level.
Following the quantitative analysis made by
\citet{Foxley-Marrable17}, in the framework of the Salpeter IMF,
only 22\% of the 650 SLSNe Ia discovered by LSST will be
standardisable due to the microlensing effects. Lensed images are
standardisable in regions of low convergence, shear and stellar
density (especially the outer image of an asymmetric double for
lenses with large $\theta_E$). Therefore, in our simulations we have
simulated two mock SLSNe Ia catalogue, 150 standardisable SLSNe Ia
without considering the microlensing effects and 500
nonstandardisable SLSNe Ia with the microlensing effect. More
specifically, we use an open-source software package for simulations
and time delay measurements of multiply imaged SNe, including an
improved characterization of the uncertainty $\Delta t$ caused by
microlensing \citep{Pierel19}. Secondly, in a system with the lensed
SNe Ia image quality typical to the HST observations, the recovery
of the relevant parameters with state-of-the-art lens modelling
techniques \citep{Suyu10,Suyu12b} make it possible to precisely
determine the lens potential. More specifically, for a single
well-measured time-delay lens system, the fractional uncertainty of
Fermat potential difference ($\Delta \phi_{i,j}$) is expected to be
determined at the level of 3\% \citep{Suyu13,Suyu14,Liao2015}.
Finally, we also include 1\% uncertainty in the lens potential to
account for the influence of the matter along the line of sight
(LOS) on strong lensing systems \citep{Jaroszyki12}, consistent with
the recent results of reconstructing the mass along a line of sight
up to intermediate redshifts \citep{panglos}.

Let us briefly describe how we simulate the unlensed SNe Ia sample.
Type Ia supernovae (SNe Ia) are particularly interesting sources
because of their nature acting as standard candles. These explosions
have nearly identical peak luminosity, which makes them excellent
distance indicators in cosmology \citep{Cao14} providing the
luminosity distances $D_L(z)$ both at lens and source redshifts:
\begin{equation}
D_L(z)=10^{(m_{X}-M_B-K_{BX})/5-5} (Mpc).
\end{equation}
where $m_X$ is the peak apparent magnitude of the supernova in
filter $X$, $M_B$ is its rest-frame B-band absolute magnitude, and
$K_{BX}$ denotes the cross-filter K-correction \citep{Kim96}. In the
catalog of lensed SNe Ia candidates, we have assumed an intrinsic
dispersion in rest-frame absolute magnitude $M_B=-19.3\pm0.2$, with
the cross-filter $K$-corrections derived from the one-component SNe
Ia spectral template \citep{Nugent02,Barbary14}. The redshift
distribution of the SNe Ia population takes the form of the
redshift-dependent SNe Ia rate \citep{Sullivan00}, which constitutes
the sampling distribution (number density) of the SNe Ia population.
In each simulation, we simulate a set of $10^5$ unlensed type-Ia
supernovae covering the redshift range of $0.0<z \leq 1.7$
\footnote{Depending on the survey strategy, LSST is expected to
yield $10^6$ type Ia supernova. However, our results show that the
impact of redshift mismatch is negligible with $10^5$ measurements
of unlensed SNe Ia, i.e., the resulting constraints will not change
with the number of unlensed SNe Ia.}. To each SNe Ia in the sample,
following the strategy described by the WFIRST Science Definition
Team (SDT) \citep{Spergel15}, we estimate the total error on the
apparent magnitude of the supernova as \citep{Hounsell17}
\begin{equation}
\sigma^2_{m_X}=\sigma^{2}_{\rm mean}+ \sigma^{2}_{\rm int} +
\sigma^{2}_{\rm lens}+ \sigma^2_{\rm sys}
\end{equation}
where the mean uncertainty is modeled as $\sigma_{\rm meas}=0.08$
mag, including both statistical measurement uncertainty and
statistical model uncertainty. The intrinsic scatter uncertainty can
be estimated with $\sigma_{\rm int} = 0.09$~mag. Another error to be
considered is $\sigma_{\rm lens}$ due to the effect of weak lensing,
and we assume $\sigma_{\rm lens} = 0.07 \times z$~mag
\citep{Holz05,Jonsson10}. Finally, the systematic uncertainty is
also considered in the SN~Ia distances, which is parameterized as
$\sigma_{sys} = 0.01(1 + z)/1.8~\rm{(mag)}$ \citep{Hounsell17}.
Denoting with $m_X$ the predicted value from our fiducial
cosmological model, we then assign to each SNe, a distance modulus
randomly generated from a Gaussian distribution centered on $m_X$
and variance $\sigma_{m_X}$ from Eq.~(12) above. A more detailed
strategy to forecast the precision on the distance modulus
determination from the SNe light curve has been described in
\citet{Cao18}.

In Table I we list the relative or absolute uncertainties of the
above mentioned factors contributing to the distance measurements.
The above simulation process is repeated 10$^3$ times, in order to
guarantee unbiased final results.

\begin{table*}
\centering
\begin{tabular}{lll}
\hline \hline
Data & Cosmic opacity (P1) & Cosmic opacity (P2) \\
\hline

SLSNe Ia (LSST; with ML)+ SNe Ia (Pantheon) & $\beta<0.020$ &$\beta<0.215$  \\
SLSNe Ia (LSST; without ML)+ SNe Ia (Pantheon) & $\beta<-0.154$ &$\beta<-0.068$  \\
SLSNe Ia (LSST)+ SNe Ia (Pantheon) & $\beta<-0.193$ &$\beta<-0.145$  \\
\hline

SLSNe Ia (LSST; with ML) + SNe Ia (LSST) & $\Delta\beta=0.075$ &$\Delta\beta=0.130$  \\
SLSNe Ia (LSST; without ML) + SNe Ia (LSST) & $\Delta\beta=0.031$ &$\Delta\beta=0.085$  \\
SLSNe Ia (LSST) + SNe Ia (LSST) & $\Delta\beta=0.027$ &$\Delta\beta=0.082$  \\
\hline
Union2.1 + Cluster \citep{Li13}&  $\beta=0.009\pm0.057$& $\beta=0.014\pm0.070$ \\
Union2.1 + $H(z)$ \citep{Liao13} & $\beta=-0.01\pm0.10$ & $\beta=-0.01\pm0.12$ \\
JLA + $H(z)$ \citep{Liao15}& $\beta=0.07\pm0.114$ & $\Box$ \\

\hline \hline
\end{tabular}
\caption{Summary of the best-fit value for the cosmic opacity
parameter obtained from different combined observations. The
Pantheon sample and simulated SNe Ia sample are respectively
combined with strongly lensed SNe Ia in a 10-year $z$-band LSST
search.} \label{comparsion_results}
\end{table*}

\section{Results and discussion}

It should be emphasized that the distance modulus of the unlensed
SNe Ia could provide the opacity-dependent time-delay distances
through Eqs.~(7)-(8). In order to avoid any bias of redshift
differences between unlensed and lensed SNe Ia, one cosmological
model-independent method is considered to associate the redshifts of
unlensed SNe Ia and lensed SNe Ia with the redshits of the lens and
source of observed from SGL systems: $|z_{SNe}-z_s|<0.005$ and
$|z_{SNe}-z_l|<0.005$. We perform Markov Chain Monte Carlo (MCMC)
minimizations to determine the cosmic-opacity parameter ($\tau$), by
minimizing the $\chi^2$ objective function defined as
\begin{equation}
\chi^2 = \sum_{1}^{i} \frac{D_{\Delta
t,i}^{lens}(z_{l,i},z_{s,i})-D_{\Delta
t,i}^{unlens}(z_{l,i},z_{s,i};
\tau)^2}{\sigma_{i,lens}^2+\sigma_{i,unlens}^2}
\end{equation}
where $D_{\Delta t,i}^{lens}$ is the time delay distance calculated
from the $ith$ strongly lensed SNe Ia (with the statistical
uncertainty $\sigma_{i,lens}$), while $D_{\Delta t,i}^{unlens}$ is
the corresponding distance inferred from the unlensed SNe Ia
observations (with the total uncertainty $\sigma_{i,unlens}$).

In order to place constraints on the cosmic opacity parameter
$\tau$, it is convenient to phenomenologically parameterize this
quantity with two monotonically increasing functions of redshift,
\begin{eqnarray}
P1. \  \tau(z)& = & 2\beta z, \\ \nonumber P2. \   \tau(z)&= &
(1+z)^{2\beta}-1. \nonumber
\end{eqnarray}
These two parameterizations, which have been widely adopted in the
literature \citep{Li13,Liao13} are basically similar for $z\ll 1$
but could differ when $z$ is not very small. One should expect the
likelihood of $\beta$ to peak at $\beta=0$, if it is consistent with
photon conservation and there is no visible violation of the
transparency of the Universe. Furthermore, we also add a prior about
the lower limit of the cosmic opacity $\beta>-0.25$, given by the
current observation of Hubble parameter ($H(z)$), the
Sunyaev-Zeldovich effect together with x-ray emission of galaxy
clusters, as well as different catalog of SNe Ia sample
\citep{Li13,Liao13,Liao15}. The graphic representations and
numerical results of the probability distribution of the opacity
parameter $\beta$ constrained from the model-independent tests are
shown in Fig.~1-4 and Table II.

\begin{figure}
\centering
\includegraphics[scale=0.4]{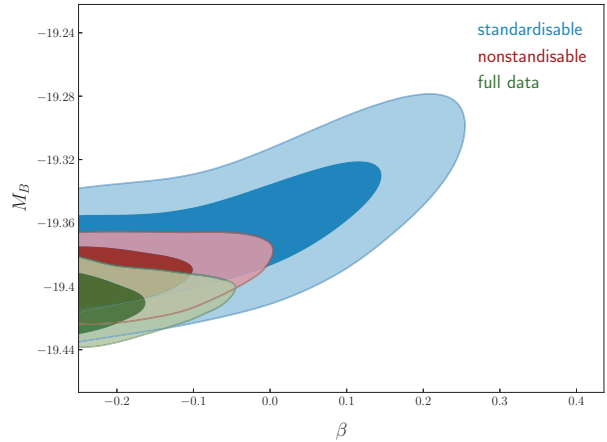}
\caption{The two-dimensional distributions of cosmic opacity
parameter $\beta$ and SNe Ia nuisance parameters ($M_B$) constrained
from the Pantheon sample in the P1 function. }\label{fig1}
\end{figure}

\begin{figure}
\centering
\includegraphics[scale=0.4]{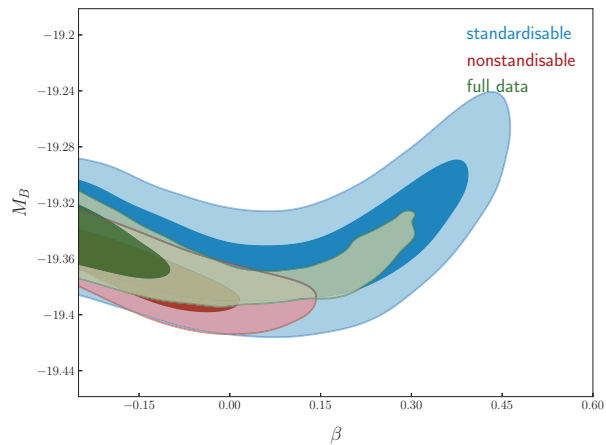}
\caption{The two-dimensional distributions of cosmic opacity
parameter $\beta$ and SNe Ia nuisance parameters ($M_B$) constrained
from the Pantheon sample in the P2 function.}\label{fig2}
\end{figure}

To get the time delay distance calculated from the unlensed SNe Ia
observations, we firstly turn to the recent Pantheon compilation by
\citet{Scolnic18} that contains 1048 SNe Ia detected by the
Pan-STARRS1 (PS1) Medium Deep Survey. Combining these SNe Ia data
with the measurements of time delay distances from lensed SNe Ia
systems, for the first $\tau(z)$ parametrization we obtain
$\beta<0.020$ and $\beta<-0.154$ (at 68.3\% confidence level) for
150 standardisable SLSNe Ia without considering the microlensing
effects, 500 nonstandardisable SLSNe Ia with the effect of
microlensing. For the full sample including 650 SLSNe Ia, the
parameter $\beta$ capturing the transparency of the Universe seems
to be vanishing: $\beta<-0.193$ for P1 function. Working on the
second $\tau(z)$ parametrization, the best-fit values are
$\beta<0.215$ and $\beta<-0.068$ for the two sub-samples
respectively including 150 standardisable SLSNe Ia without
considering the microlensing effects and 500 nonstandardisable SLSNe
Ia with the effect of microlensing. Focusing on the full sample of
strongly lensed SNe Ia observed by LSST, more stringent constraints
on the cosmic opacity will be derived: $\beta<-0.145$.
Interestingly, our findings shown in Fig.~4 illustrates the strong
degeneracy between $\beta$ and $M_B$, i.e., a lower absolute B-band
magnitude of SNe Ia will lead to a larger value of the cosmic
opacity. Such tendency, which confirms that the cosmic opacity
parameter is not independent of the SNe Ia nuisance parameters, has
also been noted and extensively discussed in the previous works
\citep{Qi19b}. Our analysis results are consistent with zero cosmic
opacity within $2\sigma$ confidence level, which indicates that
there is no significant deviation from the transparency of the
Universe at the current observational data level. However, an issue
which needs clarification is the cosmological implication of the
combination of the current Pantheon compilation and the simulated
SLSNe Ia sample. On the one hand, in order to study the systematics
and scatter in our method, we perform the diagnostics of residuals
and plot the relative residuals $(D_{\Delta t}^{obs} - D_{\Delta
t}^{th})/{\cal D}^{obs}$ as a function of $\beta$. Our results show
that there exist two different value for the cosmic opacity
parameter, $\beta=0$ and $\beta<0$, in order to achieve well
consistency between the observational time delays and their
theoretical counterparts. Therefore, our constraints on the cosmic
opacity could be biased such possibility. On the other hand, in the
framework of the current Pantheon compilation, concerning the strict
acceptable redshift difference between the unlensed SNe Ia and the
SGL system (for both the lens and source), only a limited number of
SLSNe Ia can be used to investigate the opacity of the Universe in
our cosmological-model-independent method. Therefore, in order to
draw firm and robust conclusions, one still need to minimize
uncertainties by increasing the depth and quality of observational
unlensed SNe Ia data. Earlier discussions of this issue can be found
in \citet{Cao14,Cao18,Liu19}.

\begin{figure}
\includegraphics[scale=0.5]{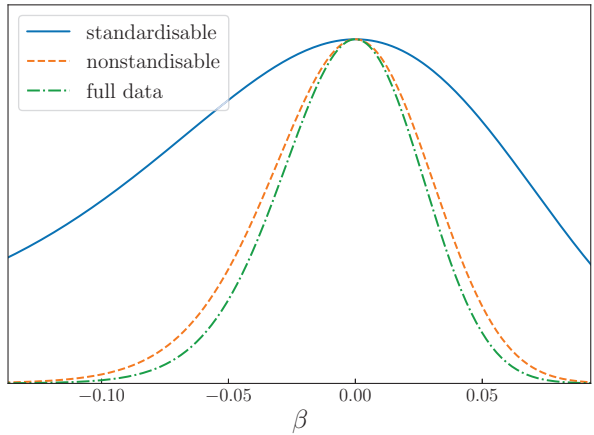}
\caption{ Probability distribution functions of opacity parameter
$\beta$ obtained from future observations of SLSNe Ia in the
forthcoming LSST survey, for the first parametrization
$\tau(z)=2\beta z$. We simulate 650 SLSNe Ia: 22\% of them would be
standarizable, the rest of them would be affected by microlensing
effects. }\label{fig3}
\end{figure}

\begin{figure}
\includegraphics[scale=0.4]{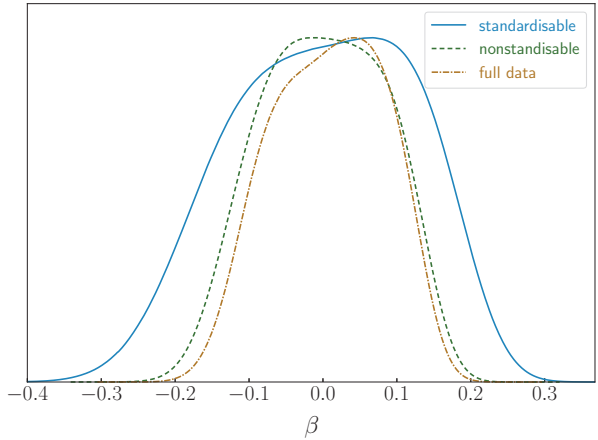}
\caption{ The same as Fig.~1, but for the second parametrization
$\tau(z)=(1+z)^{2\beta}-1$. }\label{fig4}
\end{figure}

In order to investigate the potential of future SNe Ia+SLSNe Ia to
constrain the cosmic opacity, we also derive the testing results
from simulated SNe Ia and SLSNe Ia data in Table II. The simulated
dataset (assuming a LSST-like survey) are input to the same fitting
procedure analysis we have used for the present day data. We start
our analysis with the first $\tau(z)$ parametrization modeled by
$\tau(z)= 2\beta z$, and consider two different SLSNe Ia catalogue:
150 standardisable SLSNe Ia without considering the microlensing
effects and 500 nonstandardisable LSNe Ia with the effect of
microlensing. For P1 function, the forecasts for the LSST survey is:
using only standardizable SLSNe Ia we are able to constrain the
$\beta$ parameter at the precision of $\Delta\beta=0.075$. The
remaining 78\% corrected for the microlensing effect, give
$\Delta\beta=0.031$. Finally, the full sample of 650 lensed SNe Ia
will improve the constraint to $\Delta\beta=0.027$, if the distance
measurements from unlensed counterparts are available. The results
are illustrated in Fig.~1. Such a measurement therefore provides a
stringent test of the $\beta$ parameter, and can be confidently used
to place constraints on the cosmic opacity in an opaque Universe.
Meanwhile, it is also worth investigating how the constraints depend
on the assumed $\tau(z)$ parameterization. For the P2
parametrization, the results derived from different lensed SNe Ia
sample and are shown in Fig.~3 and Table II. Working on the 150
standardisable SLSNe Ia without considering the microlensing effect,
we obtain that the opacity of the Universe could be estimated with
the precision of $\Delta\beta=0.130$ for P2 function. For the SLSNe
Ia samples including microlensing effect, the test results suggest
that the parameter $\beta$ capturing the transparency of the
Universe can be constrained with observations at the accuracy of
$\Delta\beta=0.085$. Turning to the full sample of strongly lensed
SNe Ia observed by LSST, the resulting constraint on the cosmic
opacity become $\Delta\beta=0.082$. The posterior probability
density for the $\beta$ parameter is shown in Fig.~4. From this plot
it is evident that much more severe constraints would be achieved,
and one can expect $\beta$ to be estimated with a $\Delta\beta \sim
10^{-2}$ precision. The results suggest that the tests of cosmic
opacity are not significantly sensitive to the parametrization for
$\tau(z)$. This is the most unambiguous result of the current
datasets.

It is interesting to compare our results with the previous analysis
performed to test the cosmic opacity with actual tests involving the
angular diameter distances from various astrophysical probes. In
\citet{Li13} the authors combined the two galaxy cluster samples
with luminosity distances from the Union2.1 type Ia supernova. Other
analysis were also performed in \citet{Liao13}, by fitting the
luminosity distance of SNe Ia to the observational Hubble parameter
data. Three cosmological model-independent methods (nearby SNe Ia
method, interpolation method and smoothing method) were considered
to reconstruct the opacity-free luminosity distances and associate
the redshifts of SNe Ia and $H(z)$, with the final results that an
almost transparent universe is favored. Such methodology was further
extended by \citet{Liao15}, in which type Ia supernovae observations
were considered with variable light-curve fitting parameters. The
recent determinations of the cosmic-opacity parameters from
different independent cosmological observations are also listed in
Table II. By comparing the results at 1$\sigma$, we obtain error
bars comparable or much smaller than that derived in the previous
works when the P1 and P2 functions are considered, regardless the
morphological models of galaxy clusters \citep{Li13}, the
reconstruction methods of observational Hubble parameters
\citep{Liao13,Liao15}. Therefore, the combination of strongly lensed
and unlensed SNe Ia may achieve comparable or higher precision of
the measurements of cosmic opacity than the other popular
astrophysical probes.

\section{Conclusions and discussion}

In this paper, we have discussed a new model-independent
cosmological test for the opacity of the Universe, with the
combination of the current/future measurements of type Ia supernova
sample and galactic-scale strong gravitational lensing systems with
SNe Ia acting as background sources. For the luminosity distance
$D_\mathrm{L}$ possibly affected by the cosmic opacity, we consider
the current newly-compiled SNe Ia data (Pantheon sample) and
simulated sample of SNe Ia observed by the forthcoming LSST survey,
while the observed time-delay distance $D_\mathrm{\Delta t}$
unaffected by the cosmic opacity are derived from 650 strongly
lensed SNe Ia observations in LSST. Two parameterizations,
$\tau(z)=2\beta z$ and $\tau(z)=(1+z)^{2\beta}-1$ are adopted for
the optical depth associated to the cosmic absorption.

To start with, we turn to the recent Pantheon compilation by
\citet{Scolnic18} that contains 1048 SNe Ia, all detected by the
Pan-STARRS1 (PS1) Medium Deep Survey. Combining these unlensed SNe
ia data with 650 strongly lensed SNe Ia observed by the LSST in 10
year $z$-band search, our analysis results are consistent with zero
cosmic opacity within $2\sigma$ confidence level, which indicates
that there is no significant deviation from the transparency of the
Universe at the current observational data level. Moreover, although
the tests of cosmic opacity are not significantly sensitive to its
parametrization, a degeneracy between the cosmic opacity parameter
and the absolute \textit{B}-band magnitude of SNe Ia is revealed in
this analysis. Working on more simulated unlensed SNe Ia observed by
the forthcoming LSST survey in a 10 year LSST $z$-band search, our
results show that the strongly lensed SNe Ia would produce more
robust constraints on the validity of cosmic transparency (at the
precision of $\Delta\beta=10^{-2}$). Therefore, focusing on only one
specific type of standard cosmological probe, the combination of
strongly lensed and unlensed SNe Ia may achieve considerably higher
precision of the measurements of cosmic opacity than the other
popular astrophysical probes. This is the most unambiguous result of
the current analysis.

\begin{figure}
\includegraphics[scale=0.5]{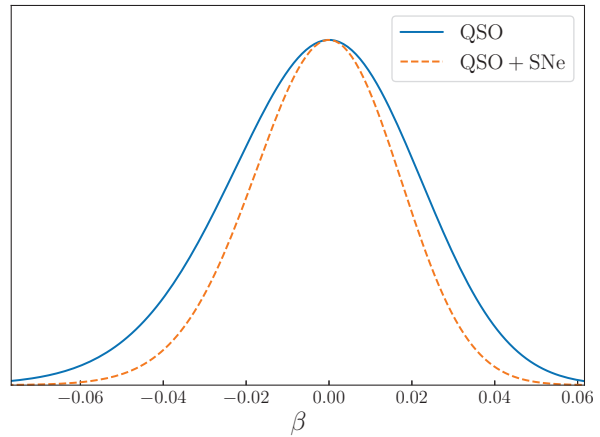}
\caption{Probability distribution of the opacity parameter $\beta$
possible to obtain with lensed quasars, as well as the combination
of lensed SNe Ia and lensed quasars (for the first parametrization
$\tau(z)=2\beta z$). }\label{fig5}
\end{figure}

There are many ways in which our methodology might be improved.
First of all, in the framework of the Chabrier IMF, 90\% lensed SNe
Ia can be classified as standard candles and insignificantly suffer
less from the microlensing effects \citep{Foxley-Marrable17}, which
makes it possible to get more precise measurements of the cosmic
opacity from future observations of strongly lensed SNe Ia.
Secondly, we may expect more vigorous and convincing constraints on
the cosmic opacity within the coming years with more precise data.
On the one hand, further progress in this direction has recently
been achieved by the very-long-baseline interferometry (VLBI)
observations, which showed that the angular diameter distances
insensitive to the opacity of the Universe can be derived from the
compact structure in intermediate luminosity radio quasars
\citep{Cao17a,Cao17b}. More importantly, LSST should detected 3000
galactic-scale strong lensing systems with quasars acting as
background sources \citep{Oguri10}, which, combined with strongly
lensed SNe Ia, will results in more stringent constraints on the
opacity of the Universe. The quasar simulation was carried out in
the following way: (I) When calculating the sampling distribution
(number density) of lensed quasars expected for the baseline survey
planned with LSST, we adopt the differential rate of lensed quasar
events as a function of $z_s$, based on the standard double power
law for the quasar luminosity function calibrated by strong lensing
effects \citep{Oguri10}. (II) In each simulation, there are 3000
lensed quasars covering the redshift range of $0.40<z<5.0$ and 1000
data points are located in the redshifts of $0.40<z<1.70$. (III)
Following the analysis of \citep{Suyu13}, the fractional
uncertainties of the Fermat potential difference and the time-delay
measurements are respectively taken at a level of 3\%. Another 1\%
uncertainty of Fermat potential reconstruction is also considered
due to LOS effects. For a good comparison, we estimate the
constraint results of the first cosmic-opacity parametrization
$\tau(z)=2\beta z$, which are specifically shown in Fig.~5.
Actually, such combination of strongly lensed SNe Ia and quasars
will enable us to get more precise measurements at the level of
$\Delta\beta=0.017$. On the other hand, the detection of
gravitational wave (GW) source with an electromagnetic counterpart
has opened an era of gravitational wave astronomy and added a new
dimension to the multi-messenger astrophysics
\citep{Abbott16,Abbott17}. Therefore, the analysis performed in this
paper can be extended to the GW domain, with the combination of
current and future available data in gravitational wave (GW) and
electromagnetic (EM) domain. One interesting approach was taken in
the paper of \citet{Liao2017}, i.e., the simultaneous detection of
strongly lensed GWs and their EM counterpart, will improve the
precision of time-delay measurements and Fermat potential
reconstruction to 0\% and 0.5\%, respectively. In this case, future
measurements of the time-delay distances of lensed gravitational
waves sources will be more competitive than the current analysis
\citep{Cao19}.

As a final remark, the method proposed in this paper, based on the
combination of strongly lensed and unlensed supernova Ia, will allow
not only to check the foundations of observational cosmology (a
transparent universe), but also open the way to identify completely
new physics (non-standard physics) if a statistically meaningful
violation of the transparent universe is observationally verified.

\section*{Acknowledgments}

This work was supported by National Key R\&D Program of China No.
2017YFA0402600; the National Natural Science Foundation of China
under Grants Nos. 11690023, 11373014, and 11633001; Beijing Talents
Fund of Organization Department of Beijing Municipal Committee of
the CPC; the Strategic Priority Research Program of the Chinese
Academy of Sciences, Grant No. XDB23000000; the Interdiscipline
Research Funds of Beijing Normal University; and the Opening Project
of Key Laboratory of Computational Astrophysics, National
Astronomical Observatories, Chinese Academy of Sciences. J.-Z. Qi
was supported by China Postdoctoral Science Foundation under Grant
No. 2017M620661, and the Fundamental Research Funds for the Central
Universities N180503014.



\begin{thebibliography}{}


\bibitem[Abbott et al.(2016)]{Abbott16} Abbott, B. P., et al. [LIGO Scientific and Virgo Collab- orations], 2016, PRL, 116, 061102
\bibitem[Abbott et al.(2017)]{Abbott17} Abbott, B. P., et al. [LIGO Scientific Collaboration, the Virgo Collaboration], 2017, PRL, 119, 161101
\bibitem[Ade et al.(2015)]{Ade15} Planck Collaboration, P. A. R. Ade et al., 2016, A\&A, 594, A13
\bibitem[Aguirre(1999)]{Aguirre99} Aguirre, A. 1999, ApJ, 525, 583
\bibitem[Allen et al.(2008)]{Allen08}  Allen, S. W., et al. 2008, MNRAS, 383, 879
\bibitem[Auger et al.(2010)]{Auger10} Auger, M. W., et al. 2010, ApJ, 724, 511
\bibitem[Avgoustidis et al.(2010)]{Avgoustidis10} Avgoustidis, A., et al. 2010, JCAP, 10, 024
\bibitem[Barbary(2014)]{Barbary14} Barbary, K. 2014, doi:10.5281/zenodo.11938
\bibitem[Bassett \& Kunz(2004)]{Bassett04} Bassett, B. A., \& Kunz, M. 2004, ApJ, 607, 661
\bibitem[Biesiada, Pi\'{o}rkowska, \& Malec(2010)]{Biesiada10} Biesiada, M., Pi\'{o}rkowska, A., \& Malec, B. 2010, MNRAS, 406, 1055
\bibitem[Bonamente et al.(2006)]{Bonamente06} Bonamente, M., et al. 2006, ApJ, 647, 25
\bibitem[Caldwell et al.(1998)]{Caldwell98}  Caldwell, R., et al. 1998, PRL, 80, 1582
\bibitem[Cao \& Liang(2011)]{Cao11} Cao, S., \& Liang, N. 2011, RAA, 11, 1199
\bibitem[Cao, et al.(2011)]{Cao11a} Cao, S., Liang, N., \& Zhu, Z.-H. 2011, MNRAS, 416, 1099
\bibitem[Cao \& Zhu(2012)]{Cao12a} Cao, S., \& Zhu, Z.-H. 20112, A\&A, 538, A43
\bibitem[Cao, Covone \& Zhu(2012)]{Cao12b} Cao, S., Covone, G., \& Zhu, Z.-H. 2012, ApJ, 755, 516
\bibitem[Cao et al.(2012)]{Cao12c} Cao, S., Pan, Y., Biesiada, M., God{\l}owski, W. \& Zhu, Z.-H. 2012, JCAP, 03, 016
\bibitem[Cao et al.(2013)]{Cao13} Cao, S., et al. 2013, RAA, 13, 15
\bibitem[Cao \& Liang(2013)]{Cao13a} Cao, S., \& Liang, N. 2013, IJMPD, 22, 1350082
\bibitem[Cao \& Zhu(2014)]{Cao14} Cao, S., \& Zhu, Z.-H. 2014, PRD, 90, 083006
\bibitem[Cao et al.(2015)]{Cao15} Cao, S., Biesiada, M., Gavazzi, R., Pi{\'o}rkowska, A. \& Zhu Z.-H. 2015, ApJ, 806, 185
\bibitem[Cao et al.(2016a)]{Cao16a} Cao, S., et al. 2016a, MNRAS, 461, 2192
\bibitem[Cao et al.(2016b)]{Cao16b} Cao, S., et al. 2016b, MNRAS, 457, 281
\bibitem[Cao et al.(2017a)]{Cao17c} Cao, S., et al. 2017a, ApJ, 835, 92
\bibitem[Cao et al.(2017b)]{Cao17a} Cao, S., Biesiada, M., Jackson, J., Zheng, X. \& Zhu Z.-H. 2017b, JCAP, 02, 012
\bibitem[Cao et al.(2017c)]{Cao17b} Cao, S., Zheng X., Biesiada M., Qi J., Chen Y. \& Zhu Z.-H. 2017c, A\&A, 606, A15
\bibitem[Cao et al.(2018)]{Cao18} Cao, S., et al. 2018, ApJ, 867, 50
\bibitem[Cao et al.(2019)]{Cao19} Cao, S., et al. 2019, Scientific Reports, in press
\bibitem[Chae(2003)]{Chae03} Chae, K.-H. 2003, MNRAS, 346, 746
\bibitem[Choi et al.(2007)]{Choi07}  Choi, Y.-Y., et al. 2007, ApJ, 658, 884
\bibitem[Collett et al.(2013)]{panglos} Collett, T. E., et al. 2013, MNRAS, 432, 679
\bibitem[Collett(2015)]{Collett15} Collett, T. E. 2015, ApJ, 811, 20
\bibitem[Collett et al.(2018)]{Collett18} Collett, T. E., et al. 2018, Science, 360, 1342
\bibitem[Corasaniti(2006)]{Corasaniti06} Corasaniti, P. S. 2006, MNRAS, 372, 191
\bibitem[Csaki et al.(2002)]{Csaki02} Csaki, C., et al. 2002, PRL, 88, 161302
\bibitem[Cullan et al.(2017)]{Cullan17} Cullan, H., et al. 2017, ApJ, 847, 128
\bibitem[Davis et al.(2007)]{Davis07}  Davis, T. M., et al. 2007, ApJ, 666, 716
\bibitem[De Filippis et al.(2005)]{Filippis05} De Filippis, E., et al. 2005, ApJ, 625, 108
\bibitem[Dobler et al.(2015)]{Greg2015} Dobler, G., et al. 2015, ApJ, 799, 168
\bibitem[Ellis(2007)]{Ellis}  Ellis, G. F. R. 2007, Gen. Rel. Grav., 39, 1047
\bibitem[Etherington(1933)]{Etherington1} Etherington, I. M. H. 1933, Phil. Mag., 15, 761
\bibitem[Etherington(2007)]{Etherington2}Etherington, I. M. H. 2007, Gen. Rel. Grav., 39, 1055
\bibitem[Foxley-Marrable et al.(2018)]{Foxley-Marrable17} Foxley-Marrable, M., et al. 2018, arXiv:1802.07738
\bibitem[Goldstein \& Nugent(2017)]{Goldstein17} Goldstein, D. A., \& Nugent, P. E. 2017, ApJL, 834, L5
\bibitem[Goobar et al.(2017)]{Goobar17} Goobar, A., et al. 2017, Science, 356, 291
\bibitem[Grillo et~al.(2008)]{Grillo08} Grillo, C., Lombardi, M., \& Bertin, G. 2008, A\&A, 477, 397
\bibitem[Holanda et al.(2012)]{Holanda12}  Holanda, R.F.L., et al. 2012, arXiv: 1207.1694
\bibitem[Holz \& Hughes(2005)]{Holz05} Holz, D. E., \& Hughes, S. A., 2005, ApJ, 629, 15
\bibitem[Hounsell et al.(2017)]{Hounsell17} Hounsell, R., et al. 2017, arXiv:1702.01747v1
\bibitem[Jaroszy\'{n}ski, et al.(2012)]{Jaroszyki12} Jaroszy\'{n}ski, M., \& Kostrzewa-Rutkowska, Z. 2012, MNRAS, 424, 325
\bibitem[J\"{o}nsson et al.(2010)]{Jonsson10} J\"{o}nsson, J., et al. 2010, MNRAS, 405, 535
\bibitem[Kessler \& Scolnic(2017)]{Kessler17} Kessler, R., \& Scolnic, D. 2017, ApJ, 836, 56
\bibitem[Kim et al.(1996)]{Kim96} Kim, A., Goobar, A., \& Perlmutter, S. 1996, PASP, 108, 190
\bibitem[Koopmans et al.(2009)]{Koopmans09} Koopmans, L. V. E., et al. 2009, ApJL, 703, L51
\bibitem[Kowalski et al.(2008)]{Kowalski08}  Kowalski, M., et al. 2008, ApJ, 686, 749
\bibitem[Li et al.(2016)]{Li16} Li, X. L., et al. 2016, RAA, 16, 84
\bibitem[Li et al.(2013)]{Li13} Li, Z., et al. 2013, PRD, 87, 103013
\bibitem[Liao et al.(2013)]{Liao13} Liao, K., et al. 2013, PLB, 718, 1166
\bibitem[Liao et al.(2015a)]{Liao15} Liao, K., et al. 2015a, PRD, 92, 123539
\bibitem[Liao et al.(2015b)]{Liao2015} Liao, K., et al. 2015b, ApJ, 800, 11
\bibitem[Liao et al.(2016)]{Liao16} Liao, K., et al. 2016, ApJ, 822, 74
\bibitem[Liao et al.(2017)]{Liao2017} Liao, K., et al. 2017, Nature Communications, 8, 1148
\bibitem[Liu et al.(2019)]{Liu19} Liu, T.-H., et al. 2019, ApJ, in press
\bibitem[Ma et al.(2017)]{Ma17} Ma, Y.-B., et al. 2017, EPJC, 77, 891
\bibitem[Ma et al.(2019)]{Ma19} Ma, Y.-B., et al. 2019, EPJC, 79, 121
\bibitem[More et al.(2009)]{More09}  More, S., et al. 2009, ApJ, 696, 1727
\bibitem[More et al.(2016)]{More16} More, S., et al. 2016, arXiv:1611.04866
\bibitem[More et al.(2017)]{More17} More, A., et al. 2017, ApJL, 835, L25
\bibitem[Marriner et al.(2011)]{Marriner11} Marriner, J., Bernstein, J. P., Kessler, R., et al., 2011, ApJ, 740, 72
\bibitem[Nugent et al.(2002)]{Nugent02} Nugent, P., Kim, A., \& Perlmutter, S. 2002, PASP, 114, 803
\bibitem[Oguri et al.(2008)]{Oguri08} Oguri, M., et al., 2008, AJ, 135, 512
\bibitem[Oguri \& Marshall(2010)]{Oguri10} Oguri, M., \& Marshall, P. J. 2010, MNRAS, 405, 2579
\bibitem[Percival et al.(2007)]{Percival07}  Percival, W. J., et al. 2007, MNRAS, 381, 1053
\bibitem[Pereira et al.(2013)]{Pereira13} Pereira, R., Thomas, R. C., Aldering, G., et al. 2013, A\&A, 554, A27
\bibitem[Perlmutter et al.(1999)]{Perlmutter99} Perlmutter, S., et al. 1999, ApJ, 517, 565
\bibitem[Pierel \& Rodney(2019)]{Pierel19} Pierel, J. D. R. \& Rodney, S. 2019, ApJ, 876, 107
\bibitem[Qi et al.(2017)]{Qi17} Qi, J. Z., et al. 2017, EPJC, 77, 02
\bibitem[Qi et al.(2018)]{Qi18} Qi, J. Z., et al. 2018, RAA, 18, 66
\bibitem[Qi et al.(2019a)]{Qi19a} Qi, J. Z., et al. 2019a, PRD, 100, 023530
\bibitem[Qi et al.(2019b)]{Qi19b} Qi, J. Z., et al. 2019b, Physics of the Dark Universe, 26, 100338
\bibitem[Ratra \& Peebles(1988)]{Ratra88}  Ratra, B., \& Peebles, P.E.J. 1988, PRD, 37, 3406
\bibitem[Refsdal(1964)]{Refsdal64} Refsdal, S. 1964, MNRAS, 128, 307
\bibitem[Riess et al.(1998)]{Riess98} Riess, A.~G., et al. 1998, AJ, 116, 1009
\bibitem[Schneider et~al.(1992)]{Schneider92} Schneider, P., Ehlers, J., \& Falco, E.~E. 1992, Gravitational Lenses
\bibitem[Scolnic et al.(2018)]{Scolnic18} Scolnic, D., et al. 2018, ApJ, 859, 101
\bibitem[Sonnenfeld et al.(2012)]{Sonnenfeld12} Sonnenfeld, A., et al. 2012, ApJ, 752, 163
\bibitem[Spergel et al.(2015)]{Spergel15} Spergel, D., et al., 2015, arXiv:1503.03757
\bibitem[Sullivan et al.(2000)]{Sullivan00} Sullivan, M., Ellis, R., Nugent, P., Smail, I., \& Madau, P. 2000, MNRAS, 319, 549
\bibitem[Suyu et al.(2010)]{Suyu10} Suyu, S. H., et al. 2010, ApJ, 711, 201
\bibitem[Suyu et al.(2012)]{Suyu12b} Suyu, S. H., et al. 2012, ApJ, 750, 10
\bibitem[Suyu et al.(2013)]{Suyu13} Suyu, S. H., et al. 2013, ApJ, 766, 70
\bibitem[Suyu et al.(2014)]{Suyu14} Suyu, S. H., et al. 2014, ApJ, 788, L35
\bibitem[Suzuki et al.(2012)]{Suzuki12} Suzuki, N., et al. 2012, ApJ, 746, 85
\bibitem[Tolman(1930)]{Tolman30} Tolman, R. C. 1930, Proc. Natl. Acad. Sci., 16, 511
\bibitem[Treu et al.(2010)]{Treu10} Treu, T. et al. 2010, ARA\&A, 48, 87
\bibitem[Walsh et al.(1979)]{Walsh79} Walsh, D., Carswell, R. F., \& Weymann, R. J. 1979, Nature, 279, 38
\bibitem[Wei(2019)]{Wei19} Wei, J.-J. 2019, arXiv:1902.00223
\bibitem[Xu et al.(2018)]{Xu18} Xu, T. P., et al. 2018, JCAP, 06, 042
\bibitem[Yang et al.(2019)]{Yang19} Yang, T., et al. 2019, Astroparticle Physics, 2019, 57
\bibitem[Young et al.(1981)]{Young81} Young, P., Gunn, J. E., Oke, J. B.,Westphal, J. A., \& Kristian, J. 1981, ApJ, 244, 736









\end{thebibliography}
\end{document}